\documentclass[aps,prd,showpacs,nofootinbib,twocolumn]{revtex4}
\usepackage{latexsym}
\usepackage{amsfonts}
\usepackage{amsbsy}
\usepackage{mathrsfs}

\begin{document}
\title{Symmetric energy-momentum tensor in Maxwell, Yang-Mills, and
Proca theories obtained using only Noether's theorem}

\date{\today}

\author{Merced Montesinos\footnote{Associate Member of the Abdus Salam
International Centre for Theoretical Physics, Trieste, Italy.}}
\email{merced@fis.cinvestav.mx} \affiliation{Departamento de F\'{\i}sica,
Cinvestav, Av. Instituto Polit\'ecnico Nacional 2508, San Pedro Zacatenco,
07360, Gustavo A. Madero, Ciudad de M\'exico, M\'exico}

\author{Ernesto Flores}
\affiliation{Facultad de F\'{\i}sica e Inteligencia Artificial, Universidad
Veracruzana,\\91000, Xalapa, Veracruz, M\'exico.}

\begin{abstract}
The symmetric and gauge-invariant energy-momentum tensors for source-free
Maxwell and Yang-Mills theories are obtained by means of translations in
spacetime via a systematic implementation of Noether's theorem. For the
source-free neutral Proca field, the same procedure yields also the symmetric
energy-momentum tensor. In all cases, the key point to get the right
expressions for the energy-momentum tensors is the appropriate handling of
their equations of motion and the Bianchi identities. It must be stressed that
these results are obtained without using Belinfante's symmetrization
techniques which are usually employed to this end.
\end{abstract}

 \maketitle

\section{Introduction}
One of the most beautiful and remarkable results in theoretical physics is
that provided by Noether's theorem, which establishes a relationship between
the symmetries of a given action and the conserved quantities for the
dynamical system associated with this action principle \cite{theo}. However,
the standard implementation of Noether's theorem to field theory leads, in the
generic case, to a non-symmetric expression for the corresponding {\it
canonical energy-momentum tensor} $\Theta^{\mu}\,_{\nu}$
\begin{eqnarray}
\Theta^{\mu}\,_{\nu} = \frac{\partial {\cal L}}{\partial (\partial_{\mu}
\phi)} \partial_{\nu} \phi  - \delta^{\mu}_{\nu} {\cal L} \, ,
\label{Canonical}
\end{eqnarray}
where $\phi$ denotes the collection of independent fields involved in the
Lagrangian density ${\cal L}$ and so in the action, and where all possible
internal indices have not been explicitly written. Next,
$\Theta^{\mu}\,_{\nu}$ is ``improved" by Belinfante's method to get the {\it
symmetric energy-momentum tensor} $T^{\mu\nu}$ \cite{Belinfante}
\begin{eqnarray}
T^{\mu\nu} = \Theta^{\mu\nu} + \partial_{\gamma} K^{\mu\nu\gamma} \, ,
\end{eqnarray}
with $K^{\mu\nu\gamma}=-K^{\mu\gamma\nu}$ and so
\begin{eqnarray}
\partial_{\nu} T^{\mu\nu}=
\partial_{\nu} \Theta^{\mu\nu}+ \partial_{\nu}\partial_{\gamma}
K^{\mu\nu\gamma}= \partial_{\nu} \Theta^{\mu\nu} \nonumber
\end{eqnarray}
(see Ref. \cite{greiner} for a detailed description of Belinfante's method). A
symmetric energy-momentum tensor, $T^{\mu\nu}$, is needed, for instance, when
such matter fields are coupled to gravity in the context of general relativity
\cite{Misner}. For Maxwell theory, taking ${\cal L}= - \frac{1}{16\pi}
F^{\mu\nu} F_{\mu\nu}$, Eq. (\ref{Canonical}) acquires the form
\begin{eqnarray}
\Theta^{\mu}\,_{\nu} =  - \frac{1}{4 \pi} F^{\mu \alpha} \partial_{\nu}
A_{\alpha} + \frac{1}{16\pi} \delta^{\mu}_{\nu} F^{\alpha\beta}
F_{\alpha\beta} \, , \label{WrongI}
\end{eqnarray}
while for Yang-Mills theory, taking ${\cal L}=- \frac{1}{16\pi} F^{\mu\nu}_a
F^a_{\mu\nu }$, Eq. (\ref{Canonical}) becomes
\begin{eqnarray}
\Theta^{\mu}\,_{\nu} = - \frac{1}{4\pi} F^{\mu \alpha}_a \partial_{\nu}
A^a_{\alpha} + \frac{1}{16 \pi} \delta^{\mu}_{\nu} F^{\alpha\beta}_a
F^a_{\alpha\beta} \, , \label{WrongII}
\end{eqnarray}
which are neither symmetric nor gauge-invariant under their corresponding
gauge transformations \cite{greiner,jackson,landau}. On the other hand, for
the neutral Proca field, taking
\begin{eqnarray}
{\cal L}= -\frac{1}{16\pi} \left( F^{\mu\nu} F_{\mu\nu} - 2 m^2 A_{\mu}
A^{\mu} \right ), \nonumber
\end{eqnarray}
Eq. (\ref{Canonical}) gives also the wrong expression \cite{greiner}
\begin{eqnarray}
\Theta^{\mu}\,_{\nu} & = & - \frac{1}{4\pi} F^{\mu\alpha} \partial_{\nu}
A_{\alpha} + \frac{1}{16\pi} \delta^{\mu}_{\nu} F^{\alpha\beta}
F_{\alpha\beta} \nonumber\\
& & \mbox{}- \frac{1}{8\pi} m^2 \delta^{\mu}_{\nu} A_{\alpha} A^{\alpha} \, .
\label{Wrong3}
\end{eqnarray}
As already mentioned, Eqs. (\ref{WrongI}), (\ref{WrongII}), and (\ref{Wrong3})
can be ``fixed" by using Belinfante's ideas \cite{Belinfante,greiner}.

Let us call the combination of standard Noether's theorem and Belinfante's
symmetrization techniques simply Noether-Belinfante's method. In spite of the
success of Belifante's symmetrization techniques to fix the canonical
energy-momentum tensor $\Theta^{\mu\nu}$ obtained by the standard Noether's
theorem, it must be emphasized that Belifante's procedure is an {\it ad hoc}
one which also has the ambiguity associated with the freedom of adding
divergence terms to $\Theta^{\mu\nu}$. In spite of these properties,
Belinfante's method has become to be very popular. Moreover, its permanent
implementation to gauge theories to ``fix" the energy-momentum tensor
$\Theta^{\mu\nu}$ has created a paradigm consisting in the claim that
Noether's theorem is {\it not} enough to determine the right form (via
spacetime translations) for the energy-momentum tensor in gauge theories.

In this paper the issue of the incompleteness or correctness of Noether's
theorem for gauge theories is analyzed, and our conclusion is that the
paradigm is not correct. Our analysis includes Maxwell, Yang-Mills, and Proca
fields, {\it i.e.}, Abelian, non-Abelian, and massive gauge fields are
analyzed. More precisely, using spacetime translations, it is shown that the
systematic implementation of Noether's theorem to the source-free Maxwell and
Yang-Mills theories leads to symmetric and gauge-invariant expressions for
their corresponding energy-momentum tensors, while for the Proca field the
procedure yields also the symmetric and right energy-momentum tensor. In all
three cases, the obtention of the right energy-momentum tensors is achieved by
taking into account both the equations of motion and the Bianchi identities
for the system under study. Therefore, the present results indicate that the
action principle has {\it all} the information required to {\it uniquely}
determine the right energy-momentum tensor and that there is {\it no} need to
use Belifante's method because the expressions obtained in this paper, which
are the right ones, follow only from a careful implementation of Noether's
theorem. This is an unexpected result which goes against the already mentioned
paradigm (for more details see Ref. \cite{flores}).

Before beginning with, some comments about our notation. Let ${\cal M}$ be the
Minkowski spacetime where the Maxwell, Yang-Mills, and Proca fields exist and
let $(x^{\mu})=(ct,x,y,z)$ be Minkowskian coordinates in it, Greek indices
$\mu,\nu...$ take the values $0,1,2,3$. The Minkowski metric is chosen to be
{\it diagonal} $(\eta_{\mu\nu})=(-1,+1,+1,+1)$. The symbol $d^4 x$ means $d
x^0 \wedge d x \wedge d y \wedge dz$ and also $\partial_{\mu} =
\frac{\partial}{\partial x^{\mu}}$. The detailed implementation of Noether's
theorem is deliberate to stress our method.

\section{Source-free Maxwell theory}
The usual action principle for source-free Maxwell theory is
\begin{eqnarray}
S[A_{\mu}] =  \alpha \int_{\cal R} d^4 x \,\, F_{\mu\nu} F^{\mu\nu} \, ,
\label{EMaction}
\end{eqnarray}
where $A=A_{\mu}(x) d x^{\mu}$, $(A_{\mu})=(\phi, {\vec A})$, is the potential
1-form and $F_{\mu\nu}=
\partial_{\mu} A_{\nu} - \partial_{\nu} A_{\mu}$ is the Faraday tensor, and
${\cal R}$ is an arbitrary region of the four-dimensional Minkowski spacetime
${\cal M}$.

The first order variation of the action (\ref{EMaction}) under the
transformation of the variables ${\tilde\delta} A_{\mu} := {\tilde A}_{\mu}
(x) - A_{\mu} (x)$ yields
\begin{eqnarray}
{\tilde\delta} S =  \int_{\cal R} d^4 x \left [ \frac{\delta S}{\delta
A_{\nu}} {\tilde\delta} A_{\nu} \right ]  + \int_{\partial {\cal R}} \left ( 4
\alpha F^{\mu\nu} {\tilde\delta} A_{\nu} \right ) d\Sigma_{\mu} \, ,
\label{FirstV}
\end{eqnarray}
with
\begin{eqnarray}
\frac{\delta S}{\delta A_{\nu}} =   - (4 \alpha
\partial_{\mu} F^{\mu\nu} ) \, , \label{FuncV}
\end{eqnarray}
and so ${\tilde\delta} S=0$ yields the equations of motion
\begin{eqnarray}
\frac{\delta S}{\delta A_{\nu}} =  - (4 \alpha
\partial_{\mu} F^{\mu\nu} ) = 0 \, , \label{MaxEq}
\end{eqnarray}
provided that the boundary term in Eq. (\ref{FirstV}) vanishes.

The action (\ref{EMaction}) is fully invariant under the Poincar\'e group
\begin{eqnarray}
S' & := & \alpha \int_{{\cal R}'} {F'}_{\mu\nu} {F'}^{\mu\nu} d^4 {x'}
\nonumber\\
& = & \alpha \int_{\cal R} F_{\mu\nu} F^{\mu\nu} d^4 x  \nonumber\\
& = & S \, .
\end{eqnarray}
The word ``fully" means that the symmetry is exact, {\it i.e.}, that the
transformed action is equal to the original action without the presence of
boundary terms. In order to apply Noether's theorem, the infinitesimal version
of the Poincar\'e transformation is needed:
\begin{eqnarray}
{x'}^{\mu} =  x^{\mu} + \delta x^{\mu}\, , \quad \delta x^{\mu} =
\varepsilon^{\mu}\,_{\nu} x^{\nu} + \varepsilon^{\mu}\, , \label{PoinT}
\end{eqnarray}
where $\varepsilon_{\mu\nu}=-\varepsilon_{\nu\mu}$ and $\varepsilon^{\mu}$ are
the infinitesimal arbitrary constant parameters associated with the
infinitesimal transformations under consideration. In addition, one has the
transformation law for the 4-potential $A_{\mu}(x)$ and the 4-gradient
$\frac{\partial}{\partial x^{\mu}}$ to first order in the parameters
\begin{eqnarray}
{A'}_{\mu} (x') & = & \frac{\partial x^{\nu}}{\partial {x'}^{\mu}} A_{\nu} (x)
= ( \delta^{\nu}\,_{\mu} -\partial_{\mu} (\delta x^{\nu}) ) A_{\nu}
\nonumber\\
& = & A_{\mu} (x) - ( \partial_{\mu} \delta x^{\nu}) A_{\nu} \, , \nonumber\\
\frac{\partial}{\partial {x'}^{\mu}} & = &   \frac{\partial x^{\nu}}{\partial
{x'}^{\mu}}  \frac{\partial}{\partial x^{\nu}}  =  \left (
\delta^{\nu}\,_{\mu}
-\partial_{\mu} (\delta x^{\nu}) \right ) \partial_{\nu} \nonumber\\
& = & \partial_{\mu} - (\partial_{\mu} \delta x^{\nu}) \partial_{\nu}\, .
\end{eqnarray}
Therefore, to first order in the parameters
\begin{eqnarray}
{F'}_{\mu\nu} (x') & := & {\partial'}_{\mu} {A'}_{\nu} - {\partial'}_{\nu}
{A'}_{\mu}\, , \nonumber\\
& = & F_{\mu\nu}(x) + ( \partial_{\nu} \delta x^{\alpha}) F_{\alpha\mu} + (
\partial_{\mu} \delta x^{\alpha} ) F_{\nu\alpha}\,\,\,\,\,\,\,\,.
\end{eqnarray}
On the other hand, one has $d^4 x' = (1+ \partial_{\mu} \delta x^{\mu}) d^4
x$. Note, however, that if one uses the explicit expression for $\delta
x^{\mu}$, then $\partial_{\mu} \delta x^{\mu}=0$ because of the antisymmetry
of $\varepsilon_{\mu\nu}$. Nevertheless, $\partial_{\mu} \delta x^{\mu}=0$
will not be used at this stage but rather at the end of the computations.
Thus, to first order
\begin{eqnarray}
S' & := & \alpha \int_{{\cal R}'} {F'}_{\mu\nu} {F'}^{\mu\nu} d^4 {x'}
\nonumber\\
& = & \alpha \int_{\cal R} F_{\mu\nu} F^{\mu\nu}
(1+ \partial_{\beta} \delta x^{\beta}) d^4 x \nonumber\\
& & + \int_{\cal R} 4 \alpha F^{\mu\nu}
(\partial_{\nu} \delta x^{\beta}) F_{\beta\mu} d^4 x  \nonumber\\
& = & S[A_{\mu}] + \alpha \int_{\cal R} F_{\mu\nu} F^{\mu\nu}
(\partial_{\beta} \delta x^{\beta}) d^4 x
\nonumber\\
& & + \int_{\cal R} \partial_{\nu} \left ( 4 \alpha F^{\mu\nu} \delta
x^{\beta}
F_{\beta\mu} \right ) d^4 x \nonumber\\
& & + \int_{\cal R} \left [ -4 \alpha \delta x^{\beta}
\partial_{\nu} ( F^{\mu\nu} F_{\beta\mu})\right ] d^4 x \, . \label{VarI}
\end{eqnarray}
To continue, it will be convenient to take into account the variation of the
action (\ref{EMaction}) with respect to the 4-potential $A_{\mu}$ and given in
Eq. (\ref{FuncV}) as well as the following definition:
\begin{eqnarray}
B_{\mu\nu\beta} :=  \partial_{\nu} F_{\beta\mu} +
\partial_{\beta} F_{\mu\nu} +
\partial_{\mu} F_{\nu\beta} \, . \label{BianchiO}
\end{eqnarray}
It is obvious that $B_{\mu\nu\beta}=0$ is equivalent to the Bianchi identities
\begin{eqnarray}
\partial_{\nu} F_{\beta\mu} + \partial_{\beta} F_{\mu\nu} +
\partial_{\mu} F_{\nu\beta} = 0. \label{BianchiIden}
\end{eqnarray}
Eqs. (\ref{MaxEq}) and (\ref{BianchiIden}) constitute the full set of Maxwell
equations. However, Eqs. (\ref{MaxEq}) and (\ref{BianchiIden}) will {\it not}
be used at this stage but rather at the end of the computations, {\it i.e.},
we will be working ``off-shell." Let $\mathscr{F}$ be the space formed by all
configurations of the gauge potentials $A=A_{\mu} (x) d x^{\mu}$, {\it i.e.},
a point in $\mathscr{F}$ is a gauge potential which does not necessarily
satisfy Eqs. (\ref{MaxEq}) and (\ref{BianchiIden}).

Going back to Eq. (\ref{VarI}), the integrand in the last term on the
right-hand side of Eq. (\ref{VarI}) can be rewritten by using Eq.
(\ref{FuncV}) as
\begin{eqnarray}
 -4 \alpha \delta x^{\beta} \partial_{\nu} ( F^{\mu\nu} F_{\beta\mu}) & = &
- 4 \alpha \delta x^{\beta} (\partial_{\nu} F^{\mu\nu}) F_{\beta\mu}
\nonumber\\
& & - 4 \alpha \delta x^{\beta} F^{\mu\nu} \partial_{\nu} F_{\beta\mu}
\nonumber\\
& = & - \delta x^{\beta} \frac{\delta S}{\delta A_{\mu}} F_{\beta\mu}
\nonumber\\
& & - 4 \alpha \delta x^{\beta} F^{\mu\nu}
\partial_{\nu} F_{\beta\mu} \, . \label{RWI}
\end{eqnarray}
Moreover, the last term on the right-hand side of Eq. (\ref{RWI}) can be
written by using the antisymmetry of $F_{\mu\nu}$ and Eq. (\ref{BianchiO}) as
\begin{eqnarray}
- 4 \alpha \delta x^{\beta} F^{\mu\nu} \partial_{\nu} F_{\beta\mu} & = & -2
\alpha \delta x^{\beta} F^{\mu\nu} \left ( \partial_{\nu}
F_{\beta\mu} - \partial_{\mu} F_{\beta\nu} \right )  \nonumber\\
& = & - 2\alpha \delta x^{\beta} F^{\mu\nu} B_{\mu\nu\beta} \nonumber\\
& & \mbox{}+ 2 \alpha \delta x^{\beta} F^{\mu\nu}
\partial_{\beta} F_{\mu\nu} \nonumber\\
& = & - 2\alpha \delta x^{\beta} F^{\mu\nu} B_{\mu\nu\beta} \nonumber\\
& & + \alpha \delta x^{\beta} \partial_{\beta} ( F^{\mu\nu} F_{\mu\nu} )
\nonumber\\
& = & - 2\alpha \delta x^{\beta} F^{\mu\nu} B_{\mu\nu\beta}
\nonumber\\
& & + \partial_{\beta} ( \alpha \delta x^{\beta} F^{\mu\nu} F_{\mu\nu}  )
\nonumber\\
& & - \alpha F^{\mu\nu} F_{\mu\nu} (\partial_{\beta} \delta x^{\beta}) \, .
\label{RWII}
\end{eqnarray}
Therefore, inserting the results of Eqs. (\ref{RWI}) and (\ref{RWII}) back
into Eq. (\ref{VarI})
\begin{eqnarray}\label{ta}
S' & = & S[A_{\mu}] + \int_{\cal R} \left [ \partial_{\beta} J^{\beta} -
\delta x^{\beta} \frac{\delta S}{\delta A_{\mu}} F_{\beta\mu} \right.
\nonumber\\
& & \left. - 2 \alpha \delta x^{\beta} F^{\mu\nu} B_{\mu\nu\beta} \right ] d^4
x \, ,
\end{eqnarray}
where
\begin{eqnarray}
J^{\beta} & := & 4 \alpha F^{\mu\beta} \delta x^{\gamma} F_{\gamma\mu} +
\alpha \delta x^{\beta} F^{\mu\nu} F_{\mu\nu} \nonumber\\
& = & T^{\beta}\,_{\gamma} \delta x^{\gamma} \, ,
\end{eqnarray}
is the Noether $4$-current and
\begin{eqnarray}\label{def}
T^{\gamma\beta} := - 4 \alpha  \left ( F^{\gamma\mu} F^{\beta}\,_{\mu} -
\frac14 \eta^{\gamma\beta} F^{\mu\nu} F_{\mu\nu} \right )\, ,
\end{eqnarray}
is the {\it energy-momentum tensor} for the electromagnetic field.

Due to the fact that the action (\ref{EMaction}) is invariant under arbitrary
transformations of the Poincar\'e group then it is, in particular, invariant
under an infinitesimal transformation and so from Eq. (\ref{ta})
\begin{eqnarray}
\int_{\cal R} \left [ \partial_{\beta} J^{\beta} - \delta x^{\beta}
\frac{\delta S}{\delta A_{\mu}} F_{\beta\mu} - 2 \alpha \delta x^{\beta}
F^{\mu\nu} B_{\mu\nu\beta}  \right ] d^4 x & = & 0 \nonumber\\
& &
\end{eqnarray}
for arbitrary spacetime regions ${\cal R}$. Therefore, the integrand in the
last equation must identically vanish
\begin{eqnarray}
\partial_{\beta} J^{\beta} =  \delta x^{\beta} \frac{\delta S}{\delta
A_{\mu}} F_{\beta\mu} + 2 \alpha \delta x^{\beta} F^{\mu\nu} B_{\mu\nu\beta}
\, . \label{NoeCon}
\end{eqnarray}
Equation (\ref{NoeCon}) is the cornerstone of the formalism of this paper.
Note that Eq. (\ref{NoeCon}) is {\it not} the so-called Noether's condition
that is usually obtained in the standard implementation of Noether's theorem.

Now, let $\overline{\mathscr{F}}$ be the phase space formed by all those
points of $\mathscr{F}$ which satisfy Eqs. (\ref{MaxEq}) and
(\ref{BianchiIden}). Therefore, for points of $\overline{\mathscr{F}}$, the
right-hand side of Eq. (\ref{NoeCon}) vanishes and the Noether $4$-current
$J^{\beta}$ is conserved
\begin{eqnarray}
\partial_{\beta} J^{\beta} =  0 \, .
\end{eqnarray}
Using the explicit form for $\delta x^{\alpha}$, the Noether 4-current
acquires the form
\begin{eqnarray}
J^{\beta} = -\frac{1}{2} \varepsilon_{\gamma\phi} M^{\beta\gamma\phi} +
\varepsilon_{\gamma} T^{\beta\gamma}\, ,
\end{eqnarray}
with
\begin{eqnarray}
M^{\beta\gamma\phi} :=  x^{\gamma} T^{\phi\beta} - x^{\phi}T^{\gamma\beta} \,
, \label{emtensor}
\end{eqnarray}
the {\it angular momentum tensor} for the electromagnetic field \cite{Misner}.
Furthermore, from the continuity equation $\partial_{\beta} J^{\beta}=0$
\begin{eqnarray}
\frac12 \varepsilon_{\gamma\phi} \left ( \partial_{\beta} M^{\beta\gamma\phi}
\right ) + \varepsilon_{\gamma} \left ( \partial_{\beta} T^{\gamma\beta}
\right ) =  0 \, ,
\end{eqnarray}
and the fact that $\varepsilon_{\gamma\phi}$ and $\varepsilon_{\gamma}$ are
independent parameters it follows that each tensor is independently conserved
\begin{eqnarray}
\partial_{\beta} M^{\beta\gamma\phi} = 0 \, , \label{ContEMI}
\end{eqnarray}
and
\begin{eqnarray}
\partial_{\beta} T^{\gamma\beta} = 0\, . \label{ContEMII}
\end{eqnarray}
Moreover, the energy-momentum tensor $ T^{\gamma\beta}$ is symmetric. This is
can be seen from its definition in Eq. (\ref{def}) or from the conservation of
the energy-momentum and angular momentum tensors, {\it i.e.}, from Eqs.
(\ref{ContEMI}) and (\ref{ContEMII}) it follows that $T^{\gamma\phi} =
T^{\phi\gamma}$. Therefore, $T^{\gamma\beta}$ is {\it conserved}, {\it
symmetric}, {\it gauge-invariant} under gauge transformations because it
depends only on $F_{\mu\nu}$, and {\it traceless} because of $T^{\mu}\,_{\mu}
= 0$. The Misner-Thorne-Wheeler's convention for the energy-momentum tensor is
\cite{Misner}
\begin{eqnarray}
T^{\mu\nu} =  \frac{1}{4\pi} \left ( F^{\mu\alpha} F^{\nu}\,_{\alpha} -
\frac{1}{4} \eta^{\mu\nu} F^{\alpha\beta} F_{\alpha\beta} \right )\, ,
\end{eqnarray}
which corresponds to set $\alpha=-1/16\pi$ and $c=1$ into the action
(\ref{EMaction}). In an explicit form the components of $T^{\mu\nu}$ read
\begin{eqnarray}
T^{00} & = & \frac{{\vec E}^2 + {\vec B}^2}{8\pi}\, , \nonumber\\
T^{0i} & = & \frac{({\vec E} \times {\vec B})^i}{4\pi}\, , \nonumber\\
T^{jk} & = & \frac{1}{4\pi} \left [ - (E^j E^k + B^j B^k)+ \frac12 ({\vec E}^2
+{\vec B}^2 )\delta^{jk} \right ]\, . \nonumber\\
& &
\end{eqnarray}

\section{Yang-Mills theory}
The reader might wonder if the procedure applied to Abelian gauge fields holds
also for non-Abelian gauge fields. The answer is in the affirmative. To see
this, the Lagrangian action for the Yang-Mills fields is considered
\cite{Yang}:
\begin{eqnarray}
S[A^a_{\mu}] = \alpha \int_{\cal R} d^4 x \,\, F^a_{\mu \nu} F^{\mu\nu}_a \, ,
\label{YMaction}
\end{eqnarray}
with
\begin{eqnarray}
F^a_{\mu\nu} = \partial_{\mu}A^a_{\nu} - \partial_{\nu}A^a_{\mu}- C^a\,_{bc}\,
A^b_{\mu} A^c_{\nu}\, ,  \label{YMstrength}
\end{eqnarray}
the strength of the Yang-Mills field $A= A^a_{\mu} d x^{\mu}\otimes T_a$ with
$T_a$ the generators of the Lie algebra of the gauge group.

Again, the first order change in the action (\ref{YMaction}) under the
transformation ${\tilde\delta} A^a_{\mu}={\tilde A}^a_{\mu}(x)- A^a_{\mu} (x)$
is:
\begin{eqnarray}
{\tilde \delta S} =  \int_{\cal R} d^4 x \left [ \frac{\delta S}{\delta
A^a_{\nu}} {\tilde\delta} A^a_{\nu}\right ] + \int_{\partial {\cal R}} \left (
4 \alpha F^{\mu\nu}_a {\tilde\delta} A^a_{\nu} \right ) d\Sigma_{\mu} \, ,
\end{eqnarray}
with\footnote{The covariant derivatives are defined as follows: $D_{\mu}
\eta^a :=
\partial_{\mu} \eta^a + C^a\,_{bc} \eta^b A^c_{\mu}$ and
$D_{\mu} \lambda_a := \partial_{\mu} \lambda_a - C^b\,_{ac} \lambda_b
A^c_{\mu}$.}
\begin{eqnarray}
\frac{\delta S}{\delta A^a_{\nu}} =  - (4 \alpha D_{\mu} F^{\mu\nu}_a ) \, ,
\label{YMFuct}
\end{eqnarray}
and so ${\tilde \delta S}=0$ gives the equations of motion
\begin{eqnarray}
D_{\mu} F^{\mu\nu}_a  = 0 \, , \label{YMEq}
\end{eqnarray}
if the boundary term in Eq. (\ref{YMEq}) vanishes.

Again, the action (\ref{EMaction}) is fully invariant under the Poincar\'e
group. In order to apply Noether's theorem, the infinitesimal version of the
Poincar\'e transformation is needed and given by Eq. (\ref{PoinT}) together
with the transformation for the Yang-Mills fields
\begin{eqnarray}
{A'}^a_{\mu} (x') & = & \frac{\partial x^{\nu}}{\partial {x'}^{\mu}} A^a_{\nu}
(x) = ( \delta^{\nu}\,_{\mu} -\partial_{\mu} (\delta x^{\nu}) ) A^a_{\nu}
\nonumber\\
& = & A^a_{\mu} (x) - ( \partial_{\mu} \delta x^{\nu}) A^a_{\nu} \, .
\end{eqnarray}
Therefore, to first order in $\delta x^{\beta}$
\begin{eqnarray}
{F'}^a_{\mu\nu} (x') & = & {\partial'}_{\mu} {A'}^a_{\nu} - {\partial'}_{\nu}
{A'}^a_{\mu} - C^a\,_{bc}\, {A'}^b_{\mu} {A'}^c_{\nu}  \nonumber\\
& = & F^a_{\mu\nu}(x) + ( \partial_{\nu} \delta x^{\alpha}) F^a_{\alpha\mu} +
( \partial_{\mu} \delta x^{\alpha} ) F^a_{\nu\alpha} \, , \quad
\end{eqnarray}
and thus to first order
\begin{eqnarray}
S' & := & \alpha \int_{{\cal R}'} {F'}^a_{\mu\nu} {F'}^{\mu\nu}_a d^4 {x'}
\nonumber\\
& = & \alpha \int_{\cal R} F^a_{\mu\nu} F^{\mu\nu}_a (1+ \partial_{\beta}
\delta x^{\beta} ) d^4 x \nonumber\\
& & + \int_{\cal R} 4 \alpha F^{\mu\nu}_a
(\partial_{\nu} \delta x^{\beta}) F^a_{\beta\mu} d^4 x  \nonumber\\
& = & S[A^a_{\mu}]  + \alpha \int_{\cal R} F^a_{\mu\nu} F^{\mu\nu}_a
(\partial_{\beta} \delta x^{\beta}) d^4 x\nonumber\\
& & + \int_{\cal R} \partial_{\nu} \left ( 4 \alpha F^{\mu\nu}_a \delta
x^{\beta}
F^a_{\beta\mu} \right ) d^4 x \nonumber\\
& & + \int_{\cal R} \left [ -4 \alpha \delta x^{\beta}
\partial_{\nu} ( F^{\mu\nu}_a F^a_{\beta\mu})\right ] d^4 x \, . \label{MMM}
\end{eqnarray}
As in the Abelian case, the object
\begin{eqnarray}
B^a_{\mu\nu\beta} =  D_{\nu} F^a_{\beta\mu} + D_{\beta} F^a_{\mu\nu} + D_{\mu}
F^a_{\nu\beta}\, , \label{XXX}
\end{eqnarray}
will be needed. The equation $B^a_{\mu\nu\beta}=0$ is equivalent to the
Bianchi identities
\begin{eqnarray}
D_{\nu} F^a_{\beta\mu} + D_{\beta} F^a_{\mu\nu} + D_{\mu} F^a_{\nu\beta} = 0
\, . \label{YYY}
\end{eqnarray}
Equations (\ref{YMEq}) and (\ref{YYY}) are the full set of Yang-Mills
equations. As in the Abelian case, we will be working `off-shell', {\it i.e.},
without using such equations at this stage but rather at the end of the
computations.

We have in hand all the elements to continue. The integrand in the last term
on the right-hand side of Eq. (\ref{MMM}) can be rewritten using Eq.
(\ref{YMFuct}) as
\begin{eqnarray}
 -4 \alpha \delta x^{\beta} \partial_{\nu} ( F^{\mu\nu}_a F^a_{\beta\mu}) & = &
- 4 \alpha \delta x^{\beta} ( D_{\nu} F^{\mu\nu}_a ) F^a_{\beta\mu}
\nonumber\\
& & - 4 \alpha \delta x^{\beta} F^{\mu\nu}_a D_{\nu} F^a_{\beta\mu}
\nonumber\\
& = & - \delta x^{\beta} \frac{\delta S}{\delta A^a_{\mu}} F^a_{\beta\mu}
\nonumber\\
& & - 4 \alpha \delta x^{\beta} F^{\mu\nu}_a D_{\nu} F^a_{\beta\mu} \, .
\label{ZZZ}
\end{eqnarray}
Rewriting the last term on the right-hand side of Eq. (\ref{ZZZ}) following
the procedure used for the Abelian case
\begin{eqnarray}
- 4 \alpha \delta x^{\beta} F^{\mu\nu}_a D_{\nu} F^a_{\beta\mu} & = &  -
2\alpha \delta x^{\beta} F^{\mu\nu}_a B^a_{\mu\nu\beta}
\nonumber\\
& & + \partial_{\beta} ( \alpha \delta x^{\beta} F^{\mu\nu}_a F^a_{\mu\nu}  )
\nonumber\\
& & - \alpha F^{\mu\nu}_a F^a_{\mu\nu} (\partial_{\beta} \delta x^{\beta}) \,
. \label{WWW}
\end{eqnarray}
Therefore, inserting the results of Eqs. (\ref{ZZZ}) and (\ref{WWW}) back into
Eq. (\ref{MMM})
\begin{eqnarray}
S' & = & S[A^a_{\mu}] + \int_{\cal R} \left [ \partial_{\beta} J^{\beta} -
\delta x^{\beta} \frac{\delta S}{\delta A^a_{\mu}} F^a_{\beta\mu}
\right. \nonumber\\
& & \left. - 2 \alpha \delta x^{\beta} F^{\mu\nu}_a B^a_{\mu\nu\beta} \right ]
d^4 x \, ,
\end{eqnarray}
with
\begin{eqnarray}
J^{\beta} & := &  4 \alpha F^{\mu\beta}_a \delta x^{\gamma} F^a_{\gamma\mu} +
\alpha \delta x^{\beta} F^{\mu\nu}_a F^a_{\mu\nu} \nonumber\\
& = & T^{\beta}\,_{\gamma} \delta x^{\gamma}\, ,
\end{eqnarray}
the Noether $4$-current and
\begin{eqnarray}\label{defYM}
T^{\gamma\beta} & :=& -4 \alpha \left ( F^{\gamma\mu}_a F^{a\beta}\,_{\mu}
-\frac14 \eta^{\gamma\beta} F^{\mu\nu}_a F^a_{\mu\nu} \right ) \, ,
\end{eqnarray}
is the {\it energy-momentum tensor} for the Yang-Mills fields.

Applying the same reasoning used in the Maxwell case, the relationship
\begin{eqnarray}
\partial_{\beta} J^{\beta} =  \delta x^{\beta} \frac{\delta S}{\delta
A^a_{\mu}} F^a_{\beta\mu} + 2 \alpha \delta x^{\beta} F^{\mu\nu}_a
B^a_{\mu\nu\beta} \, , \label{YangNoeCon}
\end{eqnarray}
is obtained. As in the Maxwell case, Eq. (\ref{YangNoeCon}) is {\it not} the
usual Noether's condition obtained by the standard Noether's theorem. As
before, Eq. (\ref{YangNoeCon}) plays a very important role here also. As
before, if the equations of motion (\ref{YMEq}) and the Bianchi identities
(\ref{YYY}) hold then the right hand side of last equation vanishes and the
Noether $4$-current $J^{\beta}$ is conserved:
\begin{eqnarray}
\partial_{\beta} J^{\beta} =  0 \, .
\end{eqnarray}
Using the explicit form for $\delta x^{\alpha}$, the Noether 4-current
acquires the form
\begin{eqnarray}
J^{\beta} =  -\frac{1}{2} \varepsilon_{\gamma\phi} M^{\beta\gamma\phi} +
\varepsilon_{\gamma} T^{\beta\gamma}\, ,
\end{eqnarray}
with
\begin{eqnarray}
M^{\beta\gamma\phi} :=  x^{\gamma} T^{\phi\beta} - x^{\phi} T^{\gamma\beta}\,
, \label{emtensorII}
\end{eqnarray}
the {\it angular momentum tensor} for the Yang-Mills fields.

Again, the same reasoning that follows Eq. (\ref{emtensor}) can be applied to
conclude that $M^{\beta\gamma\phi}$ and $T^{\gamma\beta}$ are independently
conserved and that $T^{\gamma\beta}$ is symmetric and gauge-invariant.

\section{Source-free Proca theory}
Now, it will be discussed the case of a non gauge-invariant theory: the
source-free Proca field \cite{AProca}. It is interesting to know if the
procedure of the present paper works also for this dynamical system.

The action principle for the source-free neutral Proca field is
\begin{eqnarray}
S[A_{\mu}] =  \alpha \int_{\cal R} d^4 x \left [ F_{\mu\nu} F^{\mu\nu} -2 m^2
A_{\mu} A^{\mu} \right ] \, , \label{ProcaA}
\end{eqnarray}
where $A=A_{\mu} (x) d x^{\mu}$ is the potential 1-form and
$F_{\mu\nu}=\partial_{\mu} A_{\nu} - \partial_{\nu} A_{\mu}$ its strength.

The first order variation of the action (\ref{ProcaA}) under the
transformation ${\tilde \delta} A_{\mu}= {\tilde A}_{\mu} (x) - A_{\mu} (x)$
is
\begin{eqnarray}
{\tilde \delta} S =  \int_{\cal R} d^4 x \left [ \frac{\delta S}{\delta
A_{\nu}} \right ] {\tilde\delta} A_{\nu} + \int_{\partial {\cal R}} \left ( 4
\alpha F^{\mu\nu} {\tilde\delta} A_{\nu} \right ) d \Sigma_{\mu} \, ,
\label{ProcaV}
\end{eqnarray}
with
\begin{eqnarray}
\frac{\delta S}{\delta A_{\nu}} =  - 4 \alpha \partial_{\mu} F^{\mu\nu} -
4\alpha  m^2 A^{\nu} \, , \label{NNN}
\end{eqnarray}
and so ${\tilde \delta} S=0$ yields the equations of motion
\begin{eqnarray}
\frac{\delta S}{\delta A_{\nu}} =  - 4 \alpha \partial_{\mu} F^{\mu\nu} -
4\alpha  m^2 A^{\nu} =0 \, , \label{BuBu}
\end{eqnarray}
provided that the boundary term in Eq. (\ref{ProcaV}) vanishes.

From Eq. (\ref{NNN}) it follows that
\begin{eqnarray}
\partial_{\nu} \left( \frac{\delta S}{\delta A_{\nu}} \right ) = -4 \alpha
m^2 \partial_{\nu} A^{\nu} \, , \label{KeyProca}
\end{eqnarray}
because of the antisymmetry of $F_{\mu\nu}$. Eq. (\ref{KeyProca}) will be used
in the application of Noether's theorem.

Once again, applying an infinitesimal Poincar\'e transformation
\begin{eqnarray}
S' & := & \alpha \int_{{\cal R}'} d^4 {x'} \left [ {F'}_{\mu\nu} {F'}^{\mu\nu}
- 2 m^2 {A'}_{\mu} {A'}^{\mu} \right ] \nonumber\\
& = & \alpha \int_{\cal R} d^4 x \left [ F_{\mu\nu} F^{\mu\nu} -2 m^2 A_{\mu}
A^{\mu} \right ]  (1 + \partial_{\beta} \delta x^{\beta})
\nonumber\\
& & \mbox{} + \int_{\cal R} 4 \alpha F^{\mu\nu}
(\partial_{\nu} \delta x^{\beta}) F_{\beta\mu} d^4 x  \nonumber\\
& & \mbox{} + \int_{\cal R} 4 \alpha  m^2 A^{\mu} A_{\nu} \left(
\partial_{\mu} \delta x^{\nu} \right ) d^4 x \nonumber\\
& = & S[A_{\mu}] \nonumber\\
& & \mbox{} + \alpha \int_{\cal R} d^4 x \left [ F_{\mu\nu} F^{\mu\nu} -2 m^2
A_{\mu} A^{\mu} \right ] (\partial_{\beta} \delta x^{\beta} )
\nonumber\\
& & \mbox{} + \int_{\cal R} \partial_{\nu} \left [ 4 \alpha F^{\mu\nu} \delta
x^{\beta} F_{\beta\mu} \right. \nonumber\\
&& \mbox{}+  \left. 4 \alpha  m^2 A^{\nu} A_{\mu} \delta x^{\mu}
\right ] d^4 x \nonumber\\
& & \mbox{} + \int_{\cal R} \left [ -4 \alpha \delta x^{\beta}
\partial_{\nu} ( F^{\mu\nu} F_{\beta\mu})\right ] d^4 x \nonumber\\
& & \mbox{} - \int_{\cal R} 4 \alpha  m^2 \delta x^{\nu} \partial_{\mu} \left
( A^{\mu} A_{\nu} \right ) d^4 x \, . \label{Paso}
\end{eqnarray}
By using Eq. (\ref{NNN}) the next to last term on the right-hand side of Eq.
(\ref{Paso}) acquires the form
\begin{eqnarray}
-4 \alpha \delta x^{\beta}
\partial_{\nu} ( F^{\mu\nu} F_{\beta\mu}) & = &
- 4 \alpha \delta x^{\beta} (\partial_{\nu} F^{\mu\nu}) F_{\beta\mu}
\nonumber\\
& & - 4 \alpha \delta x^{\beta} F^{\mu\nu} \partial_{\nu} F_{\beta\mu}
\nonumber\\
& = & - \delta x^{\beta} \frac{\delta S}{\delta A_{\mu}} F_{\beta\mu}
\nonumber\\
& & -4 \alpha  m^2 \delta x^{\beta} A^{\mu} F_{\beta\mu}
\nonumber\\
& & - 4 \alpha \delta x^{\beta} F^{\mu\nu}
\partial_{\nu} F_{\beta\mu} \, . \label{CuCu}
\end{eqnarray}
The last term in Eq. (\ref{CuCu}) has been already rewritten and it is given
in Eq. (\ref{RWII}). Therefore, using Eq. (\ref{RWII}), Eq. (\ref{CuCu})
becomes
\begin{eqnarray}
-4 \alpha \delta x^{\beta}
\partial_{\nu} ( F^{\mu\nu} F_{\beta\mu})
& = & - \delta x^{\beta} \frac{\delta S}{\delta A_{\mu}} F_{\beta\mu}
\nonumber\\
& & - 4 \alpha  m^2 \delta x^{\beta} A^{\mu} F_{\beta\mu}
\nonumber\\
&  & - 2\alpha \delta x^{\beta} F^{\mu\nu} B_{\mu\nu\beta}
\nonumber\\
& & + \partial_{\beta} ( \alpha \delta x^{\beta} F^{\mu\nu} F_{\mu\nu}  )
\nonumber\\
& & - \alpha F^{\mu\nu} F_{\mu\nu} (\partial_{\beta} \delta x^{\beta}) \, .
\label{pollo}
\end{eqnarray}
In a similar way, the last line in Eq. (\ref{Paso}) can be rewritten as
\begin{eqnarray}
- 4 \alpha  m^2 \delta x^{\nu} \partial_{\mu} \left ( A^{\mu} A_{\nu} \right )
& = & - 4 \alpha  m^2 \delta x^{\nu} \left ( \partial_{\mu}
A^{\mu} \right ) A_{\nu} \nonumber\\
& & - 4 \alpha  m^2 \delta x^{\nu} A^{\mu} \partial_{\mu}
A_{\nu}\nonumber\\
& = & - 4 \alpha  m^2 \delta x^{\nu} \left ( \partial_{\mu}
A^{\mu} \right ) A_{\nu} \nonumber\\
& & - 4 \alpha  m^2 \delta x^{\nu} A^{\mu} \left [ F_{\mu\nu} +
\partial_{\nu} A_{\mu} \right ] \nonumber\\
& = & \delta x^{\nu} A_{\nu} \partial_{\mu} \left ( \frac{\delta S}{\delta
A_{\mu}} \right ) \nonumber\\
& & - 4 \alpha  m^2 \delta x^{\nu} A^{\mu} F_{\mu\nu} \nonumber\\
& & - 2 \alpha  m^2 \delta x^{\nu} \partial_{\nu} \left ( A_{\mu} A^{\mu}
\right ) \nonumber\\
& = & \delta x^{\nu} A_{\nu} \partial_{\mu} \left ( \frac{\delta S}{\delta
A_{\mu}} \right ) \nonumber\\
& & + 4 \alpha m^2 \delta x^{\nu} A^{\mu} F_{\nu\mu} \nonumber\\
& & + \partial_{\nu} \left ( -2 \alpha m^2 \delta x^{\nu}  A_{\mu} A^{\mu}
\right ) \nonumber\\
& & + 2 \alpha  m^2 A_{\mu} A^{\mu} \left ( \partial_{\nu} \delta x^{\nu}
\right ) \, . \label{Pato}
\end{eqnarray}
On the right-hand side of the second equality in Eq. (\ref{Pato}), the
definition of $F_{\mu\nu}$ was used while in the third equality, Eq.
(\ref{KeyProca}) was used.

Inserting Eqs. (\ref{pollo}) and (\ref{Pato}) into Eq. (\ref{Paso})
\begin{eqnarray}
S' & = & S[A_{\mu}] + \int_{\cal R} \left [ \partial_{\beta} J^{\beta} -
\delta x^{\beta} \frac{\delta S}{\delta A_{\mu}} F_{\beta\mu}
\right. \nonumber\\
& & \left. - 2 \alpha \delta x^{\beta} F^{\mu\nu} B_{\mu\nu\beta} + \delta
x^{\nu} A_{\nu} \partial_{\mu} \left ( \frac{\delta S}{\delta A_{\mu}} \right
) \right ] d^4 x \, , \nonumber\\
& &
\end{eqnarray}
where
\begin{eqnarray}
J^{\beta} & := & 4 \alpha F^{\mu\beta} \delta x^{\gamma} F_{\gamma\mu} +
\alpha \delta x^{\beta} F^{\mu\nu} F_{\mu\nu} \nonumber\\
& & - 2 \alpha  m^2 A_{\mu} A^{\mu} \delta x^{\beta} + 4 \alpha  m^2 A^{\beta}
A_{\mu} \delta x^{\mu} \nonumber\\
& = & T^{\beta}\,_{\gamma} \delta x^{\gamma}\, ,
\end{eqnarray}
is the Noether $4$-current and
\begin{eqnarray}\label{rProca}
T^{\gamma\beta} & = & - 4 \alpha \left ( F^{\gamma\mu} F^{\beta}\,_{\mu} -
\frac14 \eta^{\gamma\beta} F^{\mu\nu} F_{\mu\nu} \right. \nonumber\\
& & \left. + \frac12 m^2 \eta^{\gamma\beta} A_{\mu} A^{\mu} -
m^2 A^{\gamma} A^{\beta} \right ) \, , \nonumber\\
\end{eqnarray}
is the {\it energy-momentum tensor} for the Proca field.

Once again, due to the fact that the action principle is invariant under the
Poincar\'e group it follows that
\begin{eqnarray}
& & \int_{\cal R} \left [ \partial_{\beta} J^{\beta} - \delta x^{\beta}
\frac{\delta S}{\delta A_{\mu}} F_{\beta\mu}
\right. \nonumber\\
& & \left. - 2 \alpha \delta x^{\beta} F^{\mu\nu} B_{\mu\nu\beta} + \delta
x^{\nu} A_{\nu} \partial_{\mu} \left ( \frac{\delta S}{\delta A_{\mu}} \right
) \right ] d^4 x =0 \, , \nonumber\\
& &
\end{eqnarray}
for arbitrary spacetime regions ${\cal R}$. Therefore, the integrand must
identically vanish:
\begin{eqnarray}
\partial_{\beta} J^{\beta} & = &   \delta x^{\beta}
\frac{\delta S}{\delta A_{\mu}} F_{\beta\mu} + 2 \alpha \delta x^{\beta}
F^{\mu\nu} B_{\mu\nu\beta} \nonumber\\
& & - \delta x^{\nu} A_{\nu}
\partial_{\mu} \left ( \frac{\delta S}{\delta A_{\mu}} \right ) \, ,
\end{eqnarray}
which is the right Noether's condition.

Therefore, if the equations of motion (\ref{BuBu}) and the Bianchi identities
hold then the Noether 4-current $J^{\beta}$ is identically conserved:
\begin{eqnarray}
\partial_{\beta} J^{\beta} =  0 \, .
\end{eqnarray}

Using the explicit expression for $\delta x^{\mu}$, the 4-current acquires the
form
\begin{eqnarray}
J^{\beta} = -\frac{1}{2} \varepsilon_{\gamma\phi} M^{\beta\gamma\phi} +
\varepsilon_{\gamma} T^{\beta\gamma} \, ,
\end{eqnarray}
with
\begin{eqnarray}
M^{\beta\gamma\phi} := x^{\gamma} T^{\phi\beta} - x^{\phi} T^{\gamma\beta} \,
,
\end{eqnarray}
the {\it angular momentum tensor}.

\section{Concluding Remarks}
It has been shown that the symmetric and gauge-invariant expressions for the
energy-momentum tensors of Maxwell and Yang-Mills fields can be obtained from
a direct implementation of Noether's theorem under a correct handling of the
terms involving the equations of motion and the Bianchi identities. The
procedure also works for the Proca fields. The reader might then wonder about
the cause of the failure of the standard Noether's approach, which leads to
Eqs. (\ref{WrongI}), (\ref{WrongII}), and (\ref{Wrong3}) instead of Eqs.
(\ref{def}), (\ref{defYM}), and (\ref{rProca}); respectively, or,
equivalently, what is then the difference between the standard Noether's
approach found in literature and the one of the present paper if after all
both approaches deal with Noether's teorem? The answer is as follows. In the
standard implementation of Noether's theorem to gauge theories only half of
the full set of equations of motion are used, the Euler-Lagrange equations.
However, when dealing with gauge theories, one has to keep in mind also the
Bianchi identities which are {\it not} taken into account in the standard
approach \cite{greiner,jackson,landau}. Nevertheless, as shown here, if they
are both taken into account, Noether's theorem yields the right expressions
for the energy-momentum tensors. Therefore, there is nothing mysterious or
wrong in the implementation of Noether's theorem to gauge theories, what
happens is just that the standard implementation of Noether's theorem is
incomplete in the sense already explained, and this is why the canonical
energy-momentum tensors so obtained need an ``improvement" via, for instance,
Belinfante's method. One could say that Noether-Belinfante's method is
equivalent to the analysis performed here in the sense that both approaches
agree in the final form for the energy-momentum tensor. This is so from an
operational (and pragmatic) viewpoint. Nevertheless, there is a key conceptual
difference between Noether-Belinfante's procedure and the one followed here.
From the viewpoint of the present paper, Belinfante's method is {\it not}
needed because the action has {\it all} the information required to {\it
uniquely} determine the right expressions for the energy-momentum tensor
$T^{\mu\nu}$ via translations in spacetime by using only Noether's theorem.
Moreover, the formalism of the present paper, in contrast to Belinfante's
method, has no ambiguities once the Lagrangian density has been chosen. In
spite of these conceptual differences, our method agrees with
Noether-Belinfante's one in the computation of the correct form for the
energy-momentum tensor. So, let us briefly say some words about the
relationship between our method and Belinfante's. As we mentioned, a key
element in our approach is the explicit incorporation and handling of the
Bianchi identities. In our opinion, the Bianchi identities (or something
equivalent to them) are ``hidden" in Belinfante's method, which allows to fix
somehow the wrong canonical energy-momentum tensor. Of course that the precise
relationship between the current formalism and the Noether-Belinfante's method
must be explained, but that is beyond the scope of this paper.

Finally, it would be interesting to generalize the results of Refs.
\cite{Nunez} and \cite{Arizmendi} to the case of gauge theories where the
Lagrangian is singular \cite{Dirac}, {\it i.e.}, $\det{\left (
\frac{\partial^2 L}{\partial {\dot q}^i \partial{\dot q}^j} \right )}=0$. Such
a generalization would involve building an action principle which would yield
the original equations of motion for the gauge system simultaneously with its
Jacobi variational equations. It is clear that the dynamical systems analyzed
in this paper could be handled in the framework of such a generalization.

\section*{Acknowledgments}
Warm thanks to G. F. Torres del Castillo, Jos\'e David Vergara, and Abdel
P\'erez-Lorenzana for their detailed reading and criticisms to the first
version of this paper. We also thank the referee for pointing out Refs.
\cite{Nunez} and \cite{Arizmendi}. This work was supported in part by the
CONACyT grant SEP-2003-C02-43939.



\begin{thebibliography}{99}
\bibitem{theo}
E. Noether, {\it Nachr. Ges. Wiss. Goettingen} {\bf 2} (1918) 235.
\bibitem{Belinfante}
J. Belinfante, {\it Physica} {\bf 6} (1939) 887; {\bf 7} (1940) 449.
\bibitem{greiner}
W. Greiner and J. Reinhardt, {\it Field Quantization} (Springer-Verlag,
Berlin, 1996).
\bibitem{Misner}
C.W. Misner, K.S. Thorne, and J.A. Wheeler, {\it Gravitation} (W. H. Freeman
and Company, New York, 1973).
\bibitem{jackson}
J.D. Jackson, {\it Classical Electrodynamics}, third edition (John Wiley and
Sons, Inc., New York, 1999).
\bibitem{landau}
L.D. Landau and E. M. Lifshitz, {\it The Classical Theory of Fields}, fourth
edition (Pergamon Press Ltd, Oxford, 1975).
\bibitem{flores}
E. Flores Gonz\'alez, Bachelor Thesis Universidad Veracruzana, M\'exico, 2006.
\bibitem{Yang}
C.N. Yang and R. Mills, {\it Phys. Rev.} {\bf 96} (1954) 191.
\bibitem{AProca}
A. Proca, {\it J. de Phys. et rad.} {\bf 7} (1936) 347; {\bf 8} (1937) 23.
\bibitem{Nunez}
H.N. N\'u\~nez-Y\'epez and A.L. Salas-Brito, {\it Phys. Lett. A} {\bf 275}
(2000) 218.
\bibitem{Arizmendi}
C.M. Arizmendi, J. Delgado, H.N. N\'u\~nez-Y\'epez, and A.L. Salas-Brito, {\it
Rev. Mex. F\'{\i}s.} {\bf 49} (2003) 298.
\bibitem{Dirac}
P.A.M. Dirac, {\it Lectures on Quantum Mechanics} (Belfer Graduate School of
Science, New York, 1964).
\end{thebibliography}
\end{document}